\documentclass[12pt]{article}
\usepackage{amssymb,amsmath,epsfig}
\allowdisplaybreaks
\begin{document}
\title{\bf Study of Some Chaotic Inflationary Models in $f(R)$ Gravity}
\author{M. Sharif \thanks{msharif.math@pu.edu.pk} and Iqra Nawazish
\thanks{iqranawazish07@gmail.com}\\
Department of Mathematics, University of the Punjab,\\
Quaid-e-Azam Campus, Lahore-54590, Pakistan.}

\date{}

\maketitle
\begin{abstract}
In this paper, we discuss inflationary scenario via scalar field and
fluid cosmology for anisotropic homogeneous universe model in $f(R)$
gravity. We consider an equation of state which corresponds to
quasi-de Sitter expansion and investigate the effect of anisotropy
parameter for different values of deviation parameter. We evaluate
potential models like linear, quadratic and quartic which correspond
to chaotic inflation. We construct the observational parameters for
power-law model of $f(R)$ gravity and construct the graphical
analysis of tensor-scalar ratio and spectral index which indicates
consistency of these parameters with Planck 2015 data.
\end{abstract}

\section{Introduction}

One of the crucial advancement on the landscape of modern cosmology
is the detection of cosmic acceleration of the universe as well as
mysteries behind its origin. The most conclusive evidence for the
present accelerated epoch appears in the measurements of supernovae
type Ia supported by some renowned observations like cosmic
microwave background (CMB), weak lensing and large scale structure.
The existence of this epoch is due to some hidden source with
surprising characteristics referred as dark energy (DE).
Astrophysical observations resolve the enigma about the birth of the
universe by introducing a model known as big-bang model (Mukhanov
2005). According to this standard model, matter or radiation
dominated phase identifies a decelerated expansion of the universe
but this decelerated expansion introduces some long standing issues
like flatness, monopole and horizon. To overcome these critical
issues, an epoch of rapid acceleration named as ``inflation'' was
suggested. It is defined as an era of few Planck lengths which
experiences a rapid exponential expansion due to some gravitational
effects (Lyth and Liddle 2009).

The idea of accelerated epoch was presented by Guth (1981) and Sato
(1981) who proposed that rapid expansion appeared due to the
existence of false vacuum filled with bubbles. This idea experienced
some shortcomings like it corresponds to de Sitter expansion and the
universe becomes inhomogeneous at the end of inflation. Such issues
lead to another version of inflation, referred as chaotic inflation
(Albrecht and Steinhardt 1982) in which a scalar field behaves like
a source of accelerated expansion. The magnitude of this scalar
field is assumed to be negatively large but the field starts rolling
down slowly towards the origin of potential. At this stage, the
potential approaches to its minimum position leading to the end of
inflation which initiates the reheating phase (Linde 1983). An
alternate approach to deal with inflationary scenario is the fluid
cosmology. It is the simplest technique which is even supported by
imperfect fluids that describe radiations and matter different from
standard one (Nojiri and Odintsov 2005).

The FRW models describe isotropic and homogeneous nature of the
universe, it ignores all structure of the universe along with
observed anisotropy in CMB temperature. Bianchi type cosmological
models are the simplest anisotropic models to analyze anisotropy
effect in the early universe on behalf of present day observations.
This anisotropy motivated many researchers to analyze inflation in
the background of anisotropic universe. For homogeneous and
anisotropic models, the anisotropy is strongly reduced by an
inflationary phase. The investigations of homogeneous and
anisotropic models also indicate that the initial anisotropy of the
universe decides the fate of the inflationary mechanisms. If the
initial anisotropy is too large then the universe cannot re-enter
into a thermal stage but for reasonably small values of anisotropy,
the inflationary phase will end with a phase transition leading to a
highly isotropic Friedmann universe (Barrow and Turner 1982). Akarsu
and Kilinc (2010) investigated Bianchi type I (BI) universe model
which describes de Sitter universe via anisotropic equation of state
(EoS) parameter. Sharif and Saleem (2014) studied locally
rotationally symmetric (LRS) BI model to analyze warm inflation
through vector fields and found consistency of this anisotropic
model with experimental data. The same authors (2015) also studied
the effects of bulk viscous pressure in warm inflation and checked
the consistency of cosmological parameters with recent WMAP7 and
Planck results.

The accelerated expansion of the universe and its evidences motivate
researchers to propose gravitational theories which can extend
general relativity to deal with puzzling nature of DE. The $f(R)$
theory is one of such modifications where $R$ represents Ricci
scalar and $f(R)$ describes a generic function. Mukhanov (2013)
analyzed cosmic inflation with a deviating EoS parameter and
formulated consistent range of observational parameters. Bamba et
al. (2014) studied reconstruction method of inflationary models and
evaluated corresponding observational parameters for different
$f(R)$ models. They found that power-law model of $f(R)$ gravity
yields most compatible results for Planck and BICEP2 constraints.
Myrzakulov and his collaborators (2015) discussed the reconstruction
technique of feasible inflationary models via scalar field and fluid
cosmology.

Artymowski and Lalak (2014) studied modified Starobinsky
inflationary model in Einstein as well as Jordan frames and found
compatible results for both BICEP2 and Planck constraints. Huang
(2014) investigated the behavior of polynomial $f(R)$ model in
inflationary paradigm and found that spectral index as well as
tensor-scalar ratio remain compatible to Planck observations. Bamba
and Odintsov (2015) discussed inflationary scenario in the
background of $f(R)$ gravity as well as loop quantum cosmology. They
concluded that for all these inflationary models, observational
parameters yield consistent results for Planck observational data.
The same authors (2016) explored inflationary universe for a viscous
fluid model and formulated observational parameters. Sharif and
Ikram (2017) explored inflationary dynamics via scalar field and
fluid cosmology of isotropic and homogeneous universe in $f(G)$
gravity. They found potential functions that correspond to chaotic
and starobinsky potential models and determined the consistent
behavior of observational parameters with Planck 2015.

The most attractive feature of chaotic inflationary model is to
describe large quantum fluctuations appearing at Planck time and
also to discuss superheavy particle production, preheating as well
as primordial gravitational waves (Kofman et al. 1994; Chung 1998).
The behavior of chaotic inflationary scenario along with
supergravity also studied on brane (Maartens et al. 2000). Gao et
al. (2014) explored chaotic inflationary model via fractional
potential and formulated observational parameters for different
fractional exponents in supergravity. Myrzakul et al. (2015) studied
chaotic inflation in higher order modified gravities via flat FRW
universe model. They investigated the behavior of massive as well as
non-massive self-interacting scalar fields and found viable
inflation for massive scalar field but obtained unrealistic
inflationary paradigm for quartic potential. We (2016, 2017a, 2017b)
have investigated the chaotic as well as warm inflationary scenario
for homogenous and isotropic flat universe model in the context of
$f(R)$ gravity.

In this paper, we study inflationary power-law model of $f(R)$
gravity using scalar field and fluid cosmology for anisotropic
homogeneous universe. The format of this paper is as follows.
Section \textbf{2} deals with some basic features of inflationary
dynamics and construct inflationary parameters. In sections
\textbf{3} and \textbf{4}, we analyze these two approaches for
different values of deviation parameter and discuss the effect of
anisotropy parameter graphically. We conclude our results in the
last section.

\section{Some Basic Features of Inflation}

We consider LRS BI universe model as
\begin{equation}\label{1}
ds^2=-\tilde{N}^2(t)dt^2+a^2(t)dx^2+b^2(t)(dy^2+dz^2),
\end{equation}
where $\tilde{N}$ represents lapse function and scale factor $a$
determines expansion of the universe along $x$-direction whereas $b$
measures the same expansion in $y$ and $z$-directions. For spatially
homogeneous metric, the normal congruence to the homogeneous
hypersurface satisfies the condition that the ratio of shear and
expansion scalars is constant which leads to a linear form,
$a=b^m,~m\neq0,1$ (Collins et al. 1980). Using this relationship,
the above model reduces to the following form
\begin{equation}\label{2}
ds^2=-\tilde{N}^2(t)dt^2+b^{2m}(t)dx^2+b^2(t)(dy^2+dz^2).
\end{equation}
The action of $f(R)$ gravity is given by (Nojiri and Odintsov 2011)
\begin{equation}\label{3}
\mathcal{A}=\int d^4x \sqrt{-g}\left(\frac{f(R)}{2\kappa^2}+
\mathcal{L}_m\right),
\end{equation}
where $\mathcal{L}_m$ is the matter Lagrangian. For perfect fluid,
the corresponding field equations become
\begin{eqnarray}\label{5}
\rho_{eff}&=&\frac{1}{2\kappa^2}\left(f-Rf_R+\frac{18(2m+1)}{(m+2)^2}H^2f_R
+6H\dot{f_R}\right),\\\nonumber
p_{eff}&=&-\frac{1}{2\kappa^2}\left(f-Rf_R+(18H^2+12\dot{H})\frac{(2m+1)}
{(m+2)^2}f_R+\frac{12(2m+1)}{(m+2)^2}H\dot{f_R}\right.
\\\label{6}&+&\left.2\ddot{f_R}\right),
\end{eqnarray}
where $R=2[(m^2+m+1)\frac{\dot{b}^2}{b^2}+(m+2)\frac{\ddot{b}}{b}],
~H(t)=\left(\frac{m+2}{3}\right)\frac{\dot{b}}{b}$,
$\rho_{eff},~p_{eff}$ represent Hubble parameter, effective energy
density and pressure, respectively. The time derivative of effective
energy density leads to
\begin{equation}\label{16}
\dot{\rho}_{eff}=\frac{1}{2\kappa^2}\left(\frac{36(2m+1)}{(m+2)^2}
H\dot{H}f_R+\frac{432(2m+1)}{(m+2)^2}H^3\dot{H}f_{RR}-288H^3\dot{H}f_{RR}\right).
\end{equation}
The Hubble flow parameters are given by
\begin{equation}\label{13}
\epsilon_1=-\frac{\dot{H}}{H^2},\quad\epsilon_2=\frac{\dot{\epsilon_1}}
{H\epsilon_1},
\end{equation}
where $\dot{H}$ is negative and $\epsilon_1,~\epsilon_2$ are
positive quantities. During inflation, $\epsilon_1$ and $\epsilon_2$
must be very small such as $\epsilon_1<<1$ and $\epsilon_2<<1$. When
$\epsilon_1=1=\epsilon_2$, the inflating universe vanishes (Linde
1990). To measure the extent of inflation, we have
\begin{equation}\label{15}
\mathcal{N}\equiv
N|_{t=t_i}=\left(\frac{m+2}{3}\right)\int^{t_f}_{t_i}
\frac{\dot{b}(t)}{b(t)}dt,
\end{equation}
where $t_f$ and $t_i$ represent cosmological time at the ending and
beginning of inflation, respectively. The approximate extent of
inflation is found to be 70 but according to fluctuation spectrum of
CMB, this limit of the e-folds becomes more smaller, i.e.,
$40<\mathcal{N}<60$. For anisotropic universe, the amplitude of
scalar and tensor power spectra
$(\Delta_\mathcal{R}^2,~\Delta_\mathcal{T}^2)$, scalar spectral
index $(n_s)$ and tensor-scalar ratio ($r$) are defined (Sharif and
Saleem 2015) as
\begin{eqnarray}\nonumber
\Delta^2_{\mathcal{R}}&=&\frac{\kappa^2H^2}{8\pi^2\epsilon_1},
\quad\Delta^2_{\mathcal{T}}=\frac{2\kappa^2H^2}{\pi^2},\quad
n_s=1-\frac{d\ln\Delta^2_\mathcal{R}}{dN},\\\label{14}
r&=&\frac{\Delta^2_\mathcal{T}}{\Delta^2_\mathcal{R}}, \quad
H=\frac{(m+2)}{3}\left(\frac{\dot{b}}{b}\right),\quad N=\mathcal{N}.
\end{eqnarray}

For general power-law model $f(R)=f_0R^n,~n\neq0,1$, where $f_0,~n$
are positive constants (Hussain et al. 2012), the field equations
are reduced to
\begin{eqnarray}\nonumber
\rho_{eff}&=&\frac{1}{2\kappa^2}\left(\frac{18(2m+1)}{(m+2)^2}
H^2nf_0R^{n-1}+6n(n-1)f_0HR^{n-2}\dot{R}
\right.\\\label{7}&+&\left.(1-n)f_0R^n\right),
\\\nonumber
p_{eff}&=&-\frac{1}{2\kappa^2}\left((1-n)f_0R^n+(18H^2+12\dot{H})\frac{(2m+1)}
{(m+2)^2}nf_0R^{n-1}+2(n-1)\right.\\\label{8}&\times&n\left.f_0R^{n-2}
\{\frac{6(2m+1)}{(m+2)^2}H\dot{R}+\ddot{R}\}+2n(n-1)(n-2)
f_0R^{n-3}\dot{R}^2\right).
\end{eqnarray}
The value of $H(t)$ and its derivative can be found using slow-roll
approximation in Eqs.(\ref{16}) and (\ref{7}) as
\begin{eqnarray}\label{17}
H^2&=&\left(\frac{4\kappa^2(m+2)^2\rho_{eff}}{12^nf_0\{2(1-n)(m+2)^2+
3n(2m+1)\}}\right)^{\frac{1}{n}},\\
\label{18} \dot{H}&=&\left(\frac{2\kappa^2(m+2)^2\dot{\rho}_{eff}H}
{12^nf_0n\{2(1-n)(m+2)^2+ 3n(2m+1)\}H^{2n}}\right).
\end{eqnarray}
Using these values in Eq.(\ref{13}), we obtain
\begin{equation}\label{19}
\epsilon_1=\frac{3(1+\omega_{eff})}{2n},
\quad\epsilon_2=-\frac{d}{dN}[\ln(1+\omega_{eff})].
\end{equation}

The effective ingredients appear due to the presence of matter
contents or scalar field. A linear relationship of these effective
quantities leads to a significant parameter, i.e., EoS parameter
($\omega_{eff}=\frac{p_{eff}}{\rho_{eff}}$) which is used to
characterize different phases of the universe. This divides DE phase
in eras like quintessence for $-1<\omega_{eff}\leq-1/3$ whereas
$\omega_{eff}<-1$ and $\omega_{eff}=-1$ correspond to phantom era
and cosmological constant (describes de Sitter expansion),
respectively. The non-vanishing accelerated expansion of the
universe is represented by these values of $\omega_{eff}$. For
vanishing rapid acceleration, there must be a small deviation such
as $\omega_{eff}\simeq-1$ instead of $\omega_{eff}=-1$. This
deviation leads to quasi-de Sitter expansion and provides a
sufficient duration of rapid expansion which elegantly admits a
graceful exit from acceleration to deceleration phase when deviating
EoS parameter approaches to the order of unity (Mukhanov 2013).

To study quasi-de Sitter inflationary epoch, we consider EoS
parameter that successfully describes the graceful exit of inflating
universe into radiation dominated era given by
\begin{equation}\label{9}
\omega_{eff}\simeq-1+\frac{\nu}{(1+N)^\mu},\quad\mu,~\nu>0,
\end{equation}
where $\nu$ is of order unity and $N$ denotes the e-folds until the
end of inflation. The corresponding conservation law
$(\dot{\rho}_{eff}+3H\rho_{eff}(1+\omega_{eff})=0)$ gives
\begin{equation}\label{10}
-\frac{d{\rho}_{eff}}{dN}+\frac{3\rho_{eff}\nu}{(N+1)^\mu}=0.
\end{equation}
Here, $\frac{d}{dt}=-H(t)\frac{d}{dN}$ and hence we obtain the
following solutions
\begin{eqnarray}\label{11}
\rho_{eff}&\simeq&\gamma(N+1)^{3\nu},\quad\mu=1,\\\label{12}
\rho_{eff}&\simeq&\gamma
\exp\left(\frac{-3\nu}{(\mu-1)(N+1)^{\mu-1}}\right),\quad\mu\neq1,
\end{eqnarray}
where $\gamma$ is the integration constant and for $N=0$,
$\rho_{eff}\simeq\gamma$ when $\mu=1$ whereas
$\rho_{eff}\simeq\gamma\exp[-3\nu/(\mu-1)]$ when $\mu\neq1$ at the
end of inflation. For Eq.(\ref{9}), the Hubble flow parameters can
be written in terms of e-folds as
\begin{equation}\label{20}
\epsilon_1\simeq\frac{3\nu}{2n(N+1)^\mu},
\quad\epsilon_2\simeq\frac{\mu}{(N+1)}.
\end{equation}
For $\mu<1$, $\epsilon_1$ is a dominant parameter whereas for
$\mu>1$, $\epsilon_2$ dominates. When $\mu=1$, both parameters play
a key role to discuss inflation at the perturbational level.

The effective energy density fluctuations are measured by the
amplitude of scalar power spectrum. The scalar power spectrum is
given by
\begin{equation}\label{21}
\Delta^2_{\mathcal{R}}=\frac{\kappa^2(m+2)^2H^2}{4\pi^2\epsilon_1
\{6(2m+1)f_R+72(2m+1)H^2f_{RR}-48(m+2)^2H^2f_{RR}\}}.
\end{equation}
Using $f(R)$ power-law model and Eq.(\ref{9}) with (\ref{11}) and
(\ref{12}), we obtain scalar power spectrum and spectral index as
\begin{eqnarray}\nonumber
\Delta^2_{\mathcal{R}}&\simeq&\gamma^{\frac{2}{n}-1}
\left(\frac{4\kappa^2(m+2)^2}{12^nf_0\{2(1-n)(m+2)^2+
3n(2m+1)\}H^{2n}}\right)^{\frac{2}{n}}\\\label{22}
&\times&\frac{(\mathcal{N}+1)^{1-3\nu+\frac{6\nu}{n}}}{4\pi^2\nu},
\quad\mu=1,
\\\nonumber\Delta^2_{\mathcal{R}}&\simeq&
\gamma^{\frac{2}{n}-1}\left(\frac{4\kappa^2(m+2)^2}
{12^nf_0\{2(1-n)(m+2)^2+
3n(2m+1)\}H^{2n}}\right)^{\frac{2}{n}}
\\\nonumber&\times&\frac{(\mathcal{N}+1)^\mu}{4\pi^2\nu}
\exp\left[\left(\frac{-3\nu}{(\mu-1)(\mathcal{N}+1)^{\mu-1}}\right)
\left(\frac{2}{n}-1\right)\right],
\\\label{23}\quad\mu&\neq&1,\\\label{24}
1-n_s&\simeq&\frac{1-3\nu+\frac{6\nu}{n}}{\mathcal{N}+1},
\quad\mu=1,\\\label{25}
\quad1-n_s&\simeq&\frac{\mu(\mathcal{N}+1)^{\mu-1}
+3\nu(\frac{2}{n}-1)}{(\mathcal{N}+1)^\mu},\quad\mu\neq1.
\end{eqnarray}
The tensor-scalar ratio for the EoS parameter (\ref{9}) is
\begin{eqnarray}\\\nonumber\quad
r&\simeq&8\kappa^2\nu\gamma^{1-\frac{1}{n}}
\left(\frac{12^nf_0\{2(1-n)(m+2)^2+3n(2m+1)\}}
{4\kappa^2(m+2)^2}\right)^{\frac{1}{n}}\\\label{26}
&\times&(\mathcal{N}+1)^{-1+3\nu-\frac{3\nu}{n}},
\quad\mu=1,\\\nonumber\quad
r&\simeq&8\kappa^2\nu\gamma^{1-\frac{1}{n}}
(\mathcal{N}+1)^{-\mu}\left(\frac{12^nf_0\{2(1-n)(m+2)^2+
3n(2m+1)\}}{4\kappa^2(m+2)^2}\right)^{\frac{1}{n}}
\\\label{27}&\times&\exp\left[\left(\frac{-3\nu}
{(\mu-1)(\mathcal{N}+1)^{\mu-1}}\right)
\left(1-\frac{1}{n}\right)\right],\quad\mu\neq1.
\end{eqnarray}
We can investigate reconstruction of different models for
$\mu=1,~\mu\neq1$. Since $n_s$ is smaller than unity in both cases
and $\epsilon_1,~\epsilon_2$ are positive, thus an elegant exit from
inflation is possible in this case. Moreover, recent observations
form Planck 2015 (Ade et al. 2016) predict the values of spectral
index and tensor-scalar ratio as $n_s=0.9666\pm0.0062$ (68\%CL) and
$r<0.10$ (95\%CL).

\section{Inflationary Model for $\mu=1$}

In this section, we reconstruct inflationary model corresponding to
spectral index (\ref{24}). The corresponding Hubble flow functions
and EoS parameter take the form
\begin{eqnarray}\label{28}
\epsilon_1\simeq\frac{3\nu}{2n(N+1)},
\quad\epsilon_2\simeq\frac{1}{(N+1)},\\\label{29}\omega_{eff}=-1
+\nu\left(\frac{\gamma}{\rho_{eff}}\right)^{\frac{1}{3\nu}}.
\end{eqnarray}
At the ending phase of inflation, $\gamma$ represents effective
energy density for $\rho_{eff}=\gamma$. The tensor-scalar ratio
turns out to be
\begin{eqnarray}\nonumber
r&\simeq&8\kappa^2\nu\gamma^{1-\frac{1}{n}}
\left(\frac{12^nf_0\{2(1-n)(m+2)^2+3n(2m+1)\}}
{4\kappa^2(m+2)^2}\right)^{\frac{1}{n}}\\\label{30}
&\times&\left(\frac{1-3\nu+\frac{6\nu}{n}}{1-n_s}
\right)^{-1+3\nu-\frac{3\nu}{n}}.
\end{eqnarray}

Now, we investigate viability of inflationary scenario in the
context of scalar field and fluid cosmology.

\subsection{Inflation via Scalar Field}

Inflation can also be analyzed by introducing a minimally coupled
scalar field $(\phi)$ subject to a potential $V(\phi)$. In this
case, Lagrangian takes the form
\begin{equation}
\mathcal{L}_{\phi}=-\frac{1}{2}g^{\alpha\beta}\partial_{\alpha}\phi
\partial_{\beta}\phi-V(\phi).
\end{equation}
The sum and difference of kinetic
$\left(\frac{\dot{\phi}^2}{2}\right)$ and potential $(V(\phi))$
energies define effective energy density $\rho_{eff}$ and pressure
$p_{eff}$, respectively which yield EoS parameter as
\begin{equation}\label{33}
\omega_{eff}=\frac{p_{eff}}{\rho_{eff}}=\frac{\frac{\dot{\phi}^2}{2}
-V(\phi)}{\frac{\dot{\phi}^2}{2}+V(\phi)}.
\end{equation}
The energy conservation law implies that
\begin{equation}\label{10'}
\ddot{\phi}+3H\dot{\phi}+V'(\phi)=0,
\end{equation}
where prime denotes derivative with respect to $\phi$. This equation
of motion is also known as scalar wave or Klein-Gordon equation.

To discuss the fluctuation patches arising from quantum fluctuations
in the early universe, chaotic inflation imposes some initial
conditions at the beginning of inflation. In chaotic inflationary
scenario, inflaton field is found to be negatively very large and
this inflationary paradigm ends for $\phi\sim M_{Pl}$. Due to this
propagating behavior of inflaton field, the corresponding chaotic
inflationary models are also known as large field models. This
inflationary scenario also describe quasi-de Sitter expansion when
$\omega_{eff}\simeq-1$ for $H=H_{dS}$ and slow-roll approximation is
valid as well. Due to slow-roll approximation, inflaton and matter
or radiation interactions are considered to be useless which implies
that kinetic energy becomes much smaller than the potential energy
of inflaton field (Guth 1981).

This approximation technique analyzes inflationary paradigm through
slow-roll parameters defined as
\begin{equation}\label{12'}
\epsilon=-\frac{\dot{H}}{H^2},
\quad\eta=-\frac{\dot{H}}{H^2}-\frac{\ddot{H}}{2H\dot{H}}
\equiv2\epsilon-\frac{\dot{\epsilon}}{2\epsilon H}.
\end{equation}
In terms of Hubble flow functions, these parameters can be expressed
as
\begin{equation}\label{36}
\epsilon=\epsilon_1,\quad\eta=2\epsilon_1-\frac{\epsilon_2}{2},
\end{equation}
which are valid for
\begin{equation}\label{13'}
\left|\frac{\dot{H}}{H^2}\right|\ll1,
\quad\left|\frac{\ddot{H}}{2H\dot{H}}\right| \ll1.
\end{equation}
In inflationary era, strong energy condition is violated which leads
to
\begin{equation}\label{11'}
\dot{\phi}^2\ll V(\phi),\quad \ddot{\phi}\ll 3H\dot{\phi}.
\end{equation}
Under the slow-roll approximation, this yields
\begin{equation}\label{14'}
3H\dot{\phi}\simeq-V'(\phi),\quad
3H\ddot{\phi}\simeq-V''(\phi)\dot{\phi}.
\end{equation}
In order to formulate inflationary model for Eq.(\ref{29}), we
obtain a relationship between kinetic energy and potential of the
field using Eq.(\ref{33}) as
\begin{equation}
\dot{\phi}\simeq\frac{\sqrt{\nu}\gamma^{\frac{1}{6\nu}}}
{V(\phi)^{\frac{1-3\nu}{6\nu}}}.
\end{equation}
For the potential, we take the first equation of (\ref{14'}) which
yields
\begin{eqnarray}\nonumber
V(\phi)&=&\left[3\sqrt{\nu}\gamma^{\frac{1}{6\nu}}
\left(\frac{4\kappa^2(m+2)^2}{12^nf_0\{2(1-n)(m+2)^2+
3n(2m+1)\}}\right)^{\frac{1}{2n}}\right]^{\frac{6n\nu}
{n+3n\nu-3\nu}}\\\label{38}&\times&\left(\frac{n+3n\nu-3\nu}
{6n\nu}\right)^{\frac{6n\nu}{n+3n\nu-3\nu}}
(-\phi)^{\frac{6n\nu}{n+3n\nu-3\nu}}.
\end{eqnarray}

For $\nu=\frac{1}{6}$, the EoS parameter becomes
\begin{equation}\label{44}
\omega_{eff}=-1 +\frac{1}{6}\left(\frac{\gamma}{\rho_{eff}}\right),
\end{equation}
and the corresponding inflaton takes the form
\begin{equation*}
\phi=\phi_i+\sqrt{\frac{\gamma}{6}}(t-t_i),
\end{equation*}
where $\phi_i$ is the integration constant. This corresponds to the
negatively large field at initial phase of inflation. For such
scalar field, the potential and Hubble function become
\begin{figure}
\epsfig{file=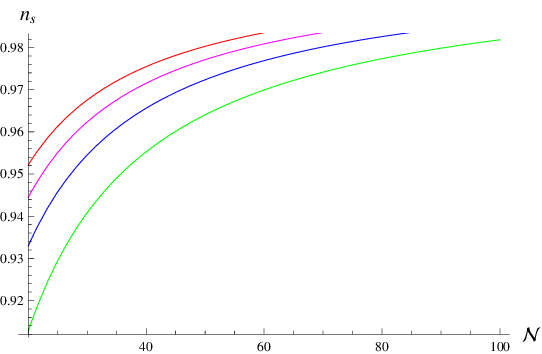, width=0.5\linewidth}\epsfig{file=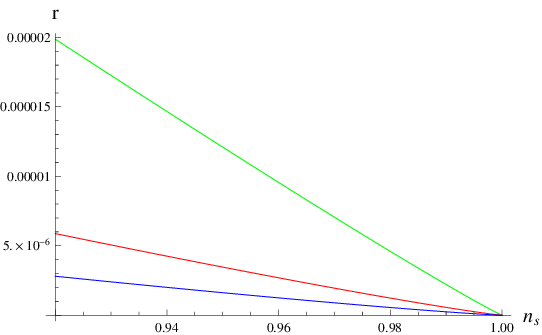,
width=0.5\linewidth}\caption{$n_{s}$ versus $\mathcal{N}$ (left) for
$n=0.75$ (green), $n=1.1$ (blue), $n=1.5$ (magenta) and $n=1.98$
(red) whereas $r$ versus $n_{s}$ (right) for $n=0.9,~m=0.3$ (green),
$n=0.8,~m=0.5$ (blue) and $n=0.75,~m=0.8$ (red).}
\end{figure}
\begin{eqnarray*}
V(\phi)&=&\frac{\chi_1}{(\frac{2n}{3n-1})}(-\phi)^{\frac{2n}{3n-1}},
\\\nonumber H(t)&=&\left[\left(\frac{\chi_1(3n-1)}{2n}\right)
\left(\frac{4\kappa^2(m+2)^2}{12^nf_0\{2(1-n)(m+2)^2+
3n(2m+1)\}}\right)\right]^{\frac{1}{2n}}\\\nonumber&\times&(-\phi)^{\frac{1}
{3n-1}},
\end{eqnarray*}
where $\chi_1$ is constant given as
\begin{eqnarray}\nonumber
\chi_1&=&\left[\left\{\sqrt{\frac{3}{2}}\gamma
\left(\frac{4\kappa^2(m+2)^2}{12^nf_0\{2(1-n)(m+2)^2+
3n(2m+1)\}}\right)^{\frac{1}{2n}}\right\}^{2n}\right.\\\nonumber
&\times&\left.\left(\frac{3n-1}{2n}\right)^{1-n}\right]^{\frac{1}{3n-1}}.
\end{eqnarray}
Notice that $V(\phi)$ is linear for $n=1$. The slow-roll parameters,
spectral index and tensor-scalar ratio are
\begin{eqnarray}\label{43}
\epsilon&=&\frac{1}{4n(N+1)},\quad\eta=2(1-n)\epsilon,\quad
n_s=1-\frac{n+2}{2n(\mathcal{N}+1)},\\\nonumber\quad
r&=&\frac{4\kappa^2}{3}\gamma^{1-\frac{1}{n}}
\left(\frac{12^nf_0\{2(1-n)(m+2)^2+3n(2m+1)\}}
{4\kappa^2(m+2)^2}\right)^{\frac{1}{n}}\\\label{30}
&\times&\left(\frac{n+2}{2n(1-n_s)}\right)^{-\frac{1+n}{2n}}.
\end{eqnarray}
In Figure \textbf{1}, the left plot indicates that the e-folds start
decreasing as $n$ increases whereas the right panel shows that
tensor-scalar ratio is compatible for all considered values of $n$
and $m$. The consistent behavior of $r$ is shown in both plots of
Figure \textbf{2}.
\begin{figure}
\epsfig{file=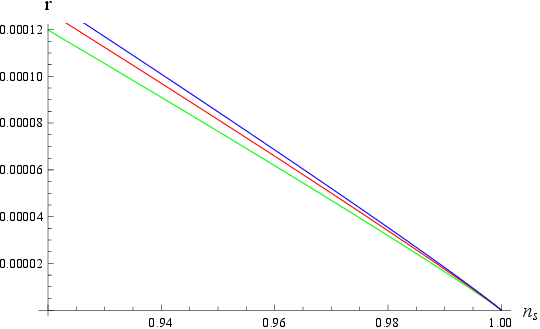, width=0.5\linewidth}\epsfig{file=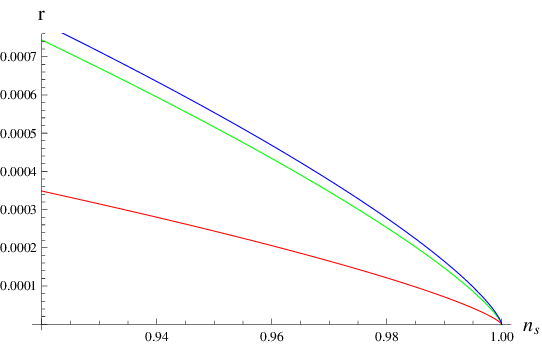,
width=0.5\linewidth}\caption{$r$ versus $n_{s}$ (left)$n=1.1,~m=0.3$
(green), $n=1.1,~m=0.5$ (blue) and $n=1.1,~m=0.8$ (red) while $r$
versus $n_{s}$ (right) for $n=1.8,~m=0.3$ (green), $n=1.92,~m=0.5$
(blue) and $n=1.98,~m=0.8$ (red).}
\end{figure}

When $\nu=\frac{1}{3}$, Eq.(\ref{29}) becomes
\begin{equation}\label{45}
\omega_{eff}=-1 +\frac{1}{3}\left(\frac{\gamma}{\rho_{eff}}\right).
\end{equation}
At the beginning of inflation, this effective parameter leads to the
following from of inflaton field, potential and Hubble function as
\begin{eqnarray*}
\phi&=&\phi_i+\sqrt{\frac{\gamma}{3}}(t-t_i),\quad
V(\phi)=\chi_2(1-\frac{1}{2n})(-\phi)^{\frac{1}{1-\frac{1}{2n}}},
\\\nonumber H(t)&=&\left[\chi_2\left(1-\frac{1}{2n}\right)
\left(\frac{4\kappa^2(m+2)^2}{12^nf_0\{2(1-n)(m+2)^2+
3n(2m+1)\}}\right)\right]^{\frac{1}{2n}}\\\nonumber&\times&(-\phi)^{\frac{1}
{2n(1-\frac{1}{2n})}},
\end{eqnarray*}
where
\begin{equation}\nonumber
\chi_2=\left[\left(1-\frac{1}{2n}\right)^{\frac{1}{2n}}(6\gamma)^{\frac{1}{2}}
\left(\frac{4\kappa^2(m+2)^2}{12^nf_0\{2(1-n)(m+2)^2+
3n(2m+1)\}}\right)^{\frac{1}{2n}}\right]^{\frac{1}{1-\frac{1}{2n}}},
\end{equation}
and $V(\phi)$ called a generalized quadratic potential of chaotic
inflation. The corresponding slow-roll parameters, spectral index
and tensor-scalar ratio are
\begin{eqnarray}\label{43}
\epsilon&=&\frac{1}{2n(N+1)},\quad\eta=(2-n)\epsilon,\quad
n_s=1-\frac{2}{n(\mathcal{N}+1)},\\\nonumber\quad
r&=&\frac{8\kappa^2}{3}\gamma^{1-\frac{1}{n}}
\left(\frac{12^nf_0\{2(1-n)(m+2)^2+3n(2m+1)\}}
{4\kappa^2(m+2)^2}\right)^{\frac{1}{n}}\\\label{30}
&\times&\left(\frac{2}{n(1-n_s)}\right)^{-\frac{1}{n}}.
\end{eqnarray}
The left graph in Figure \textbf{3} describes that for quadratic
potential, the e-folds are getting smaller as $n$ gets larger. The
right plot shows that $r$ is consistent with Planck constraint
whether we increase or decrease the value of anisotropy parameter
whereas Figure \textbf{4} indicates the same results for $r$.
\begin{figure}
\epsfig{file=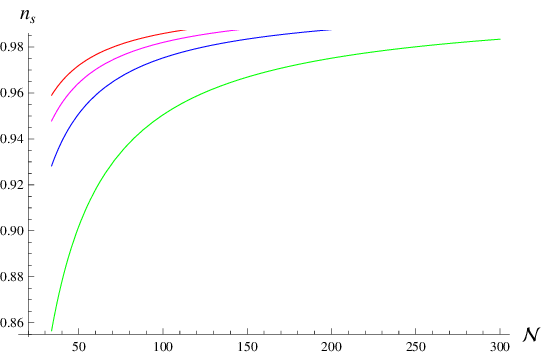, width=0.5\linewidth}\epsfig{file=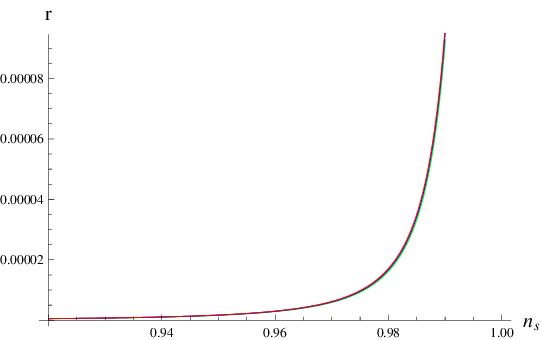,
width=0.5\linewidth}\caption{$n_{s}$ versus $\mathcal{N}$ (left) for
$n=0.4$ (green), $n=0.8$ (blue), $n=1.1$ (magenta) and $n=1.4$ (red)
and $r$ versus $n_{s}$ (right) for $n=0.4,~m=0.3$ (green),
$n=0.4,~m=0.5$ (blue) and $n=0.4,~m=0.8$ (red).}
\end{figure}

For $\nu=\frac{2}{3}$, we have
\begin{equation}\label{46}
\omega_{eff}=-1 +\frac{2}{3}\left(\frac{\gamma}{\rho_{eff}}\right).
\end{equation}
The scalar field and $H(t)$ can be expressed as
\begin{eqnarray*}
\phi&=&\phi_i+\sqrt{\frac{2\gamma}{3}}(t-t_i),\quad
V(\phi)=\frac{\chi_3}{\frac{4n}{3n-2}}(-\phi)^{\frac{4n}{3n-2}},
\\\nonumber H(t)&=&\left[\frac{\chi_3(3n-2)}{4n}
\left(\frac{4\kappa^2(m+2)^2}{12^nf_0\{2(1-n)(m+2)^2+
3n(2m+1)\}}\right)\right]^{\frac{1}{2n}}\\\nonumber&\times&(-\phi)^{\frac{2}
{3n-2}}.
\end{eqnarray*}
The potential of the scalar field corresponds to quartic potential
model of chaotic inflation for $n=1$ with coupling constant given as
\begin{equation}\nonumber
\chi_3=\left[\sqrt{6}\left(\frac{3n-2}{4n}\right)^{n+2}\gamma^{\frac{1}{4}}
\left(\frac{4\kappa^2(m+2)^2}{12^nf_0\{2(1-n)(m+2)^2+
3n(2m+1)\}}\right)^{\frac{1}{2n}}\right]^{\frac{4n}{3n-2}}.
\end{equation}
In this case, the slow-roll and observational parameters become
\begin{figure}
\epsfig{file=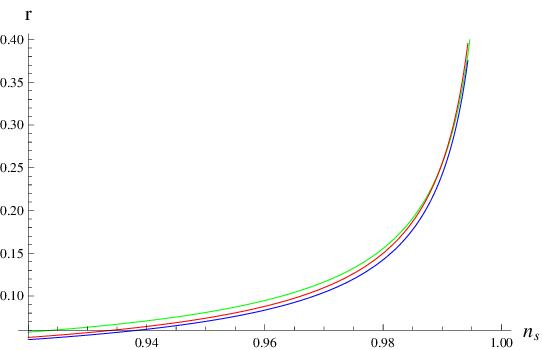, width=0.5\linewidth}\epsfig{file=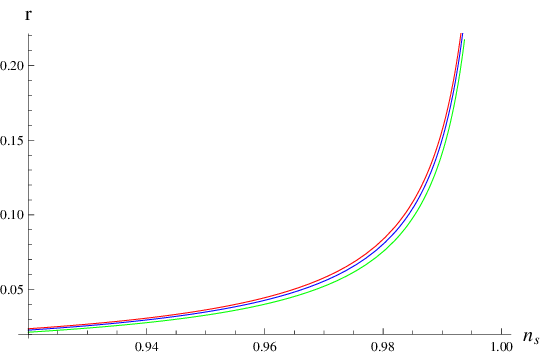,
width=0.5\linewidth}\caption{$r$ versus $n_{s}$ (left)$n=1.1,~m=0.3$
(green), $n=1.1,~m=0.5$ (blue) and $n=1.1,~m=0.8$ (red) while $r$
versus $n_{s}$ (right) for $n=1.4,~m=0.3$ (green), $n=1.3,~m=0.5$
(blue) and $n=1.3,~m=0.8$ (red).}
\end{figure}
\begin{eqnarray}\label{ss}
\epsilon&=&\frac{1}{n(N+1)},\quad\eta=\frac{(4-n)\epsilon}{2},\quad
n_s=1-\frac{4-n}{n(\mathcal{N}+1)},\\\nonumber\quad
r&=&\frac{16\kappa^2}{3}\gamma^{1-\frac{1}{n}}
\left(\frac{12^nf_0\{2(1-n)(m+2)^2+3n(2m+1)\}}
{4\kappa^2(m+2)^2}\right)^{\frac{1}{n}}\\\label{tt}
&\times&\left(\frac{4-n}{n(1-n_s)}\right)^{-1+\frac{2}{n}}.
\end{eqnarray}
In Figure \textbf{5}, the left panel represents that there exists an
inverse relation between $n$ and $\mathcal{N}$, i.e., e-folds
decreases when $n$ increases and vice-versa. The best fit value of
the e-folds is obtained for $n=1.5$ with quartic potential. The
right plot indicates that we obtain a consistent range for
$m=0.3,~0.5$ and $0.8$ whereas $n$ remains the same, i.e., $n=1.1$.
Figures \textbf{6} and \textbf{7} yield a compatible range of
tensor-scalar ratio for different values of $m$ and $n$.
\begin{figure}
\epsfig{file=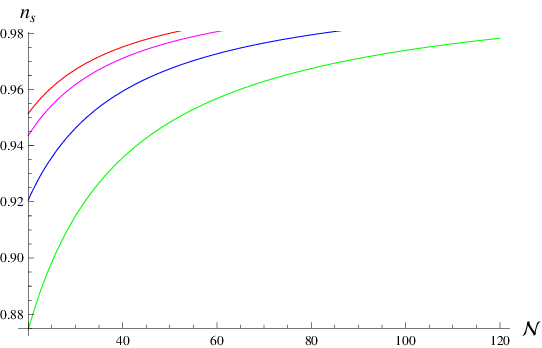, width=0.5\linewidth}\epsfig{file=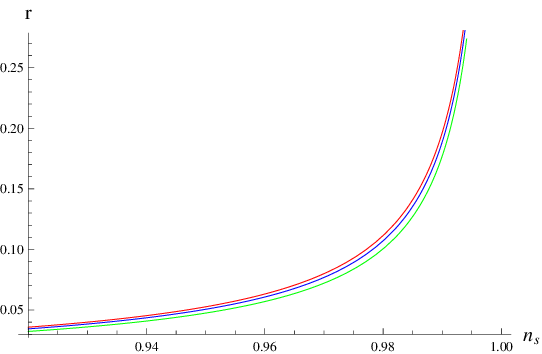,
width=0.5\linewidth}\caption{$n_{s}$ versus $\mathcal{N}$ (left) for
$n=1.1$ (green), $n=1.5$ (blue), $n=1.83$ (magenta) and $n=1.98$
(red) and $r$ versus $n_{s}$ (right) for $n=1.1,~m=0.3$ (green),
$n=1.1,~m=0.5$ (blue) and $n=1.1,~m=0.8$ (red).}
\end{figure}
\begin{figure}
\epsfig{file=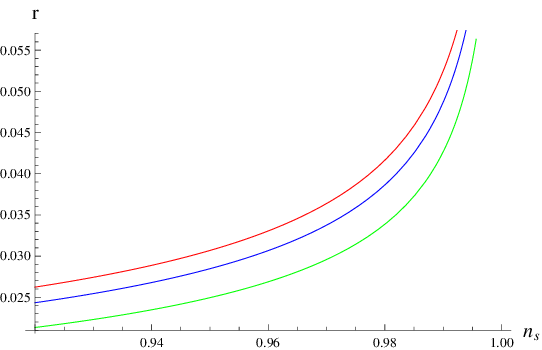, width=0.5\linewidth}\epsfig{file=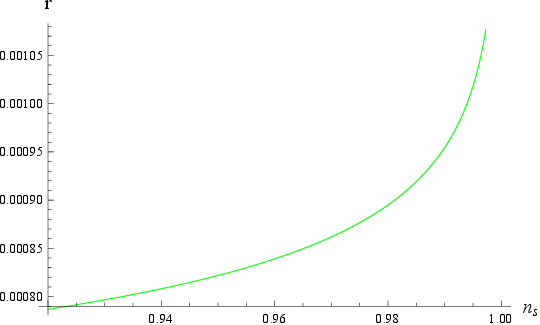,
width=0.5\linewidth}\caption{$r$ versus $n_s$ (left) for $n=1.5$,
$m=0.3$ (green), $m=0.5$ (blue), $m=0.8$ (red) whereas $r$ versus
$n_{s}$ (right) for $n=1.83,~m=0.3$ (green).}
\end{figure}
\begin{figure}
\epsfig{file=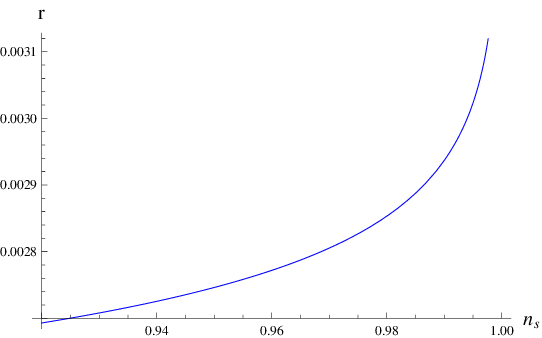, width=0.5\linewidth}\epsfig{file=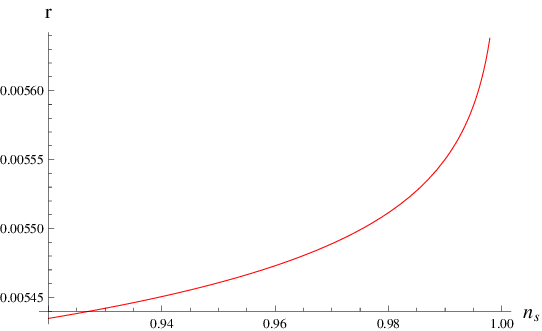,
width=0.5\linewidth}\caption{$r$ versus $n_s$ (left) for
$n=1.92,~m=0.5$ (blue) while $r$ versus $n_{s}$ (right) for
$n=1.98,~m=0.8$ (red).}
\end{figure}

\subsection{Inflation via Fluid Cosmology}

This is a well-known approach to any cosmological phenomenon which
deals with perfect as well as imperfect fluid corresponding to
ordinary radiation or matter in the universe. A straight forward
description of rapid and uniform accelerated expansion of the
universe is given by exotic matter which is governed by EoS
different from radiation or ordinary matter. To discuss a graceful
exit of rapid acceleration into a deceleration phase, quasi-de
Sitter expansion in which EoS parameter depends on energy density.
We consider the EoS parameter as
\begin{equation}\label{47}
\omega(\tilde{\rho})=-1+\nu\left(\frac{\gamma}{\tilde{\rho}}
\right)^{\frac{1}{3\nu}},
\end{equation}
where $\tilde{\rho}$ and $\tilde{p}$ represent energy density and
pressure of inhomogeneous fluid. The energy density from
conservation law and Hubble parameter for $\nu=\frac{1}{6}$ are
\begin{eqnarray}\nonumber
\tilde{\rho}&=&\left[\mu_0-\gamma^2\left(\frac{4n-1}{12n}\right)
\left(\frac{4\kappa^2(m+2)^2}{12^nf_0\{2(1-n)(m+2)^2+
3n(2m+1)\}}\right)^{\frac{1}{2n}}\right.\\\label{48}&\times&\left.
(t-t_i)\right]^{\frac{2n}{4n-1}},\\\nonumber
H&=&\left(\frac{4\kappa^2(m+2)^2}{12^nf_0\{2(1-n)(m+2)^2+
3n(2m+1)\}}\right)^{\frac{1}{2n}}\left[\mu_0-\gamma^2\left(\frac{4n-1}{12n}\right)
\right.\\\label{49}&\times&\left.\left(\frac{4\kappa^2(m+2)^2}{12^nf_0\{2(1-n)(m+2)^2+
3n(2m+1)\}}\right)^{\frac{1}{2n}}(t-t_i)\right]^{\frac{1}{4n-1}},
\end{eqnarray}
where $\mu_0$ is an integration constant of the quasi-de Sitter
expansion. Inflation occurs when $t$ approaches to $t_i$ for which
$\epsilon_1$ and $\epsilon_2$ become
\begin{equation*}
\epsilon_1=\frac{1}{4n}\left(\frac{\gamma}{\tilde\rho}\right)^2,
\quad\epsilon_2=\left(\frac{\gamma}{\tilde\rho}\right)^2.
\end{equation*}

These parameters recover the expressions of slow-roll parameters,
spectral index and tensor-scalar ratio when $\nu=\frac{1}{6}$ for
the scalar field. Equations (\ref{11}) and (\ref{47}) lead to a
relationship between number of e-folds and energy density of
inhomogeneous fluid as
\begin{equation*}
N+1=\left(\frac{\tilde\rho}{\gamma}\right)^2.
\end{equation*}
This provides a condition for the ending of inflation, i.e.,
$(\tilde\rho=\gamma)$ and scale factor takes the form
\begin{equation*}
a(t)=a_f\exp\left[1-\left(\frac{\tilde\rho}{\gamma}\right)^2\right],
\end{equation*}
where $a_f$ denotes the scale factor at the end of inflation. For
$\nu=\frac{1}{3}$, the energy density and Hubble parameter become
\begin{eqnarray}\nonumber
\tilde{\rho}&=&\left[\mu_0-\gamma\left(\frac{2n-1}{2n}\right)
\left(\frac{4\kappa^2(m+2)^2}{12^nf_0\{2(1-n)(m+2)^2+
3n(2m+1)\}}\right)^{\frac{1}{2n}}\right.\\\label{48'}&\times&\left.
(t-t_i)\right]^{\frac{2n}{2n-1}},\\\nonumber
H&=&\left(\frac{4\kappa^2(m+2)^2}{12^nf_0\{2(1-n)(m+2)^2+
3n(2m+1)\}}\right)^{\frac{1}{2n}}\left[\mu_0-\gamma\left(\frac{2n-1}{2n}\right)
\right.\\\label{49'}&\times&\left.\left(\frac{4\kappa^2(m+2)^2}{12^nf_0\{2(1-n)(m+2)^2+
3n(2m+1)\}}\right)^{\frac{1}{2n}}(t-t_i)\right]^{\frac{1}{2n-1}}.
\end{eqnarray}
Using Eqs.(\ref{48'}) and (\ref{49'}), we obtain Hubble flow
functions at $t=t_i$ as
\begin{equation*}
\epsilon_1=\frac{1}{2n}\left(\frac{\gamma}{\tilde\rho}\right),
\quad\epsilon_2=\left(\frac{\gamma}{\tilde\rho}\right).
\end{equation*}
The resulting slow-roll parameters, spectral index and tensor-scalar
ratio turn out to be the same as for $\nu=\frac{1}{3}$ in the scalar
field. The scale factor becomes
\begin{equation*}
a(t)=a_f\exp\left[1-\left(\frac{\tilde\rho}{\gamma}\right)\right],
\end{equation*}
where $N+1=\frac{\tilde\rho}{\gamma}$. When $\nu=\frac{2}{3}$, the
energy density, Hubble parameter and its flow functions for
inhomogeneous fluid are
\begin{eqnarray}\nonumber
\tilde{\rho}&=&\left[\mu_0-\gamma^{\frac{1}{2}}\left(\frac{n-1}{n}\right)
\left(\frac{4\kappa^2(m+2)^2}{12^nf_0\{2(1-n)(m+2)^2+
3n(2m+1)\}}\right)^{\frac{1}{2n}}\right.\\\label{48''}&\times&\left.
(t-t_i)\right]^{\frac{2n}{n-1}},\\\nonumber
H&=&\left(\frac{4\kappa^2(m+2)^2}{12^nf_0\{2(1-n)(m+2)^2+
3n(2m+1)\}}\right)^{\frac{1}{2n}}\left[\mu_0-\gamma^
{\frac{1}{2}}\left(\frac{n-1}{n}\right)
\right.\\\label{49''}&\times&\left.\left(\frac{4\kappa^2(m+2)^2}
{12^nf_0\{2(1-n)(m+2)^2+3n(2m+1)\}}\right)^
{\frac{1}{2n}}(t-t_i)\right]^{\frac{1}{n-1}},\\\nonumber
\epsilon_1&=&\frac{1}{n}\left(\frac{\gamma}{\tilde\rho}\right),
\quad\epsilon_2=\left(\frac{\gamma}{\tilde\rho}\right)^{\frac{1}{2}}.
\end{eqnarray}
The parameters $\epsilon_1$ and $\epsilon_2$ recover the slow-roll
and observational parameters formulated for $\nu=\frac{2}{3}$ with
scalar field.

Finally, we take EoS for $\nu=\frac{1}{3}$ and use the field
equations (\ref{5}) and (\ref{6}) which yield
\begin{equation}\label{66}
2\ddot{f}_{R}-6H\dot{f}_{R}\left[1-2\left(\frac{2m+1}{(m+2)^{2}}
\right)\right]+12\dot{H}\left(\frac{2m+1}{(m+2)^{2}}\right)
=-\frac{2\kappa^{2}\gamma}{3}.
\end{equation}
The Hubble parameter and its derivative take the form
\begin{equation*}
H=\sqrt{\frac{\kappa^{2}\gamma(m+2)^{2}(N+1)}{9(2m+1)}},\quad
\dot{H}=\frac{\kappa^{2}\gamma(m+2)^{2}}{18(2m+1)},
\end{equation*}
which can also expressed in terms of e-folds at the end of inflation
as
\begin{equation*}
\frac{dH}{dN}=\frac{1}{2}\sqrt{\frac{\kappa^{2}\gamma(m+2)^{2}}
{9(2m+1)(N+1)}}.
\end{equation*}
Thus, we can reconstruct $f(R)$ model by inserting the above
derivative in Eq.(\ref{66}) leading to
\begin{equation*}
\frac{a(N+1)}{3}\ddot{f}_{R}+a\dot{f}_{R}\left[\frac{1}{6}+(N+1)
\left(1-\frac{2}{a}\right)\right]-f_{R}=-1,
\end{equation*}
where $a=\frac{(m+2)^2}{2m+1}$. Integrating the above equation, we
obtain
\begin{eqnarray}\nonumber
f_{R}(R)&=&\sqrt{N+1}\exp\left(-\frac{3(a-2)N}{a}\right)\left[c_1
U\left(-\frac{1-a}{a-2},\frac{3}{2},\frac{3(a-2)(N+1)}{a}\right)\right.
\\\label{67}&+&\left.c_2L_{\frac{1-a}{a-2}}^{\frac{1}{2}}
\left(\frac{3(a-2)(N+1)}{a}\right)\right]+1.
\end{eqnarray}
Here, $c_1$ and $c_2$ are integration constants whereas $U$
represents confluent hypergeometric function and $L$ denotes
associated Laguerre polynomial. This equation (\ref{67}) can also be
expressed in terms of $R$ by taking
\begin{equation}
R=\kappa^2\gamma(4N+3),\quad
N=\frac{3}{4}\left(\frac{(2m+1)R}{(m+2)^2\kappa^2\gamma}-1\right),
\end{equation}
which implies that
\begin{eqnarray}\label{68}
f_{R}(R)&=&1+\frac{1}{2}\sqrt{\frac{3R}{a\kappa^{2}\gamma}+1}
\exp\left(\frac{-9(a-2)}{4a}\left(\frac{R}{a\kappa^{2}\gamma}
-1\right)\right)\\\nonumber&\times&\left[c_{1}U
\left(-\frac{1-a}{a-2},\frac{3}{2},
\frac{9(a-2)}{4a}\left(\frac{R}{\kappa^{2}a\gamma}+1\right)\right)
\right.\\\nonumber&+&\left.c_{2}L^{\frac{1}{2}}_{\frac{1-a}{a-2}}
\left(\frac{9(a-2)}{4a}\left(\frac{R}
{a\kappa^{2}\gamma+1}\right)\right)\right].
\end{eqnarray}
Its integration leads to
\begin{eqnarray}\nonumber
f(R)&=&R+\frac{1}{2}\int\sqrt{\frac{3R}{a\kappa^{2}\gamma}+1}
\exp\left(\frac{-9(a-2)}{4a}\left(\frac{R}{a\kappa^{2}\gamma}
-1\right)\right)\\\nonumber&\times&\left[c_{1}
U\left(-\frac{1-a}{a-2},\frac{3}{2},
\frac{9(a-2)}{4a}\left(\frac{R}{\kappa^{2}a\gamma}+1\right)\right)
\right.\\\nonumber&+&\left.c_{2}
L^{\frac{1}{2}}_{\frac{1-a}{a-2}}\left(\frac{9(a-2)}{4a}
\left(\frac{R}{a\kappa^{2}\gamma+1}\right)\right)\right]+c_3,
\end{eqnarray}
where $c_3$ is an integration constant. This $f(R)$ model
corresponds to Starobinsky inflationary model for $n,~m,~f_0=1$.

\section{Inflationary Model for $\mu\neq1$}

Here, we would like to investigate the existence of the viable
inflationary models for $\mu=2$ with spectral index and
tensor-scalar ratio given in Eqs.(\ref{25}) and (\ref{27}). In this
case, the EoS (\ref{9}) and Hubble flow functions (\ref{20}) reduce
to
\begin{eqnarray}\label{53}
\omega_{eff}&=&-1+\left(\frac{1}{9\nu}\right)\log^2
\left[\frac{\rho_{eff}}{p_{eff}}\right],\\\label{54}
\epsilon_1&=&\frac{3\nu}{2n(N+1)^2},\quad\epsilon_2=\frac{2}{(N+1)},
\end{eqnarray}
where $\epsilon_2$ is much larger than $\epsilon_1$. The
observational parameters $(n_s,~r)$ become
\begin{eqnarray}\label{55}
n_s&=&1-\frac{2}{(\mathcal{N}+1)},\\\nonumber
r&=&2\kappa^2\nu\gamma^{1-\frac{1}{2n}}(1-n_s)^2
\left(\frac{12^nf_0\{2(1-n)(m+2)^2+3n(2m+1)\}}{4\kappa^2(m+2)^2}\right)
^{\frac{1}{n}}\\\label{56}&\times&\exp\left[-\frac{3\nu(1-n_s)(n-1)}{2n}\right],
\end{eqnarray}
where $(\mathcal{N}+1)^2=\frac{4}{(1-n_s)^2}$. In this case,
tensor-scalar ratio is found to be inconsistent with Planck
constraint whereas $\mathcal{N}=59$.

For the inflation via scalar field, the kinetic energy and potential
function for $n=1,~2,~3$ are formulated using Eqs.(\ref{33}),
(\ref{14'}) and (\ref{53}) as
\begin{eqnarray}
\dot{\phi}&\simeq&\frac{1}{3}\sqrt{\frac{V(\phi)}{\nu}}
\left(\frac{\gamma}{V(\phi)-1}\right),\\\nonumber
V(\phi)&=&2\gamma'\left\{c_4-\sqrt{\frac{2}{\nu}}
\left(\frac{4\kappa^2(m+2)^2}{12f_0\{3(2m+1)\}}
\right)^{\frac{1}{2}}\phi\right\},\\\nonumber
V(\phi)&=&\left[\frac{5\gamma'}{2} \left\{c_4-\sqrt{\frac{2}{\nu}}
\left(\frac{4\kappa^2(m+2)^2}{144f_0\{-2(m+2)^2+6(2m+1)\}}
\right)^{\frac{1}{4}}\phi\right\}\right]^{\frac{4}{5}},\\\nonumber
V(\phi)&=&\left[\frac{8\gamma'}{3}\left\{c_4-\sqrt{\frac{2}{\nu}}
\left(\frac{4\kappa^2(m+2)^2}{12^3f_0\{-4(m+2)^2+9(2m+1)\}}
\right)^{\frac{1}{6}}\phi\right\}\right]^{\frac{3}{4}},
\end{eqnarray}
where $c_4$ and $\gamma'$ are integration constants. The above forms
of potentials define massless large inflaton field in terms of
fractional potential models that can be generalized as
\begin{eqnarray*}
&&\left\{c_4-\sqrt{\frac{2}{\nu}}
\left(\frac{4\kappa^2(m+2)^2}{12^nf_0\{2(1-n)(m+2)^2+3n(2m+1)\}}
\right)^{\frac{1}{2n}}\phi\right\}\frac{\gamma'(3n-1)}{n}\\\nonumber&&=
V^{\frac{3n-1}{2n}}(\phi)\left\{1+\left(\frac{3n-1}{5n-1}\right)
\frac{V(\phi)}{\gamma'}\right\}.
\end{eqnarray*}
The fractional potential function corresponds to Starobinsky model
for $m,~n=1$ (Starobinsky 1980). The slow-roll and observational
parameters for $c_4=2,~\nu=\frac{1}{2}$ become
\begin{eqnarray}\label{59}
\epsilon&=&\frac{3}{4n(N+1)^2},\quad\eta=-\frac{1}{(N+1)}+\frac{3}{2n(N+1)^2},
\\\label{61}\quad n_s&=&1-\frac{2}{\mathcal{N}+1},\\\nonumber
r&=&\kappa^2\gamma^{1-\frac{1}{2n}}(1-n_s)^2
\left(\frac{12^nf_0\{2(1-n)(m+2)^2+3n(2m+1)\}}
{4\kappa^2(m+2)^2}\right)^{\frac{1}{n}}\\\label{60}
&\times&\exp\left[-\frac{3(n-1)(1-n_s)}{4n}\right].
\end{eqnarray}

In order to discuss inflation via fluid cosmology, we take
inhomogeneous fluid so that EoS takes the form
\begin{equation}
\omega(\tilde\rho)=-1+\frac{1}{9\nu}\log^2
\left(\frac{\tilde\rho}{\gamma}\right).
\end{equation}
Inserting Eq.(\ref{17}) in the conservation law, we obtain
\begin{eqnarray}\nonumber
\tilde\rho&=&\gamma\left[1-\left(\frac{12^nf_0\{2(1-n)(m+2)^2+3n(2m+1)\}}
{4\kappa^2(m+2)^2}\right)^{\frac{1}{2n}}\frac{3\nu}
{\gamma^{\frac{1}{2n}}(t_f-t)}\right],\\\label{62}\\\nonumber
H&=&\left[1-3\nu \left(\frac{12^nf_0\{2(1-n)(m+2)^2+3n(2m+1)\}}
{4\kappa^2(m+2)^2}\right)^{\frac{1}{2n}}\frac{1}
{\gamma^{\frac{1}{2n}}(t_f-t)}\right]^{\frac{1}{2n}}\\\label{63}&\times&\left(\frac
{4\kappa^2(m+2)^2}{12^nf_0\{2(1-n)(m+2)^2+3n(2m+1)\}}\right)^
{\frac{1}{2n}}\gamma^{\frac{1}{2n}}.
\end{eqnarray}
In this limit $t<<t_e$, the Hubble parameter become
\begin{eqnarray}\label{64}
\epsilon_1&=&\frac{3\nu}{2n\gamma^{\frac{1}{n}}(t_f-t)^2}
\left(\frac{12^nf_0\{2(1-n)(m+2)^2+3n(2m+1)\}}
{4\kappa^2(m+2)^2}\right)^{\frac{1}{n}},\\\label{65}
\epsilon_2&=&\frac{2}{\gamma^{\frac{1}{2n}}(t_f-t)}
\left(\frac{12^nf_0\{2(1-n)(m+2)^2+3n(2m+1)\}}
{4\kappa^2(m+2)^2}\right)^{\frac{1}{2n}}.
\end{eqnarray}
To describe the duration of inflation, the number of e-folds and
scale factor turn out to be
\begin{eqnarray}\nonumber
N&=&\left(\frac{4\kappa^2(m+2)^2\gamma}{12^nf_0\{2(1-n)(m+2)^2+3n(2m+1)\}}
\right)^{\frac{1}{2n}}(t_f-t)-1,\\\nonumber
a(t)&=&a_f\exp\left[1-\left(\frac{4\gamma\kappa^2(m+2)^2}
{12^nf_0\{2(1-n)(m+2)^2+3n(2m+1)\}}\right)^{\frac{1}{2n}}(t-t_f)\right].
\end{eqnarray}
The parameters $\epsilon_1$ and $\epsilon_2$ recover the expressions
given in Eqs.(\ref{54})-(\ref{56}).

\section{Concluding Remarks}

This paper is devoted to study inflation via two approaches scalar
field and fluid cosmology in $f(R)$ gravity using LRS BI universe
model. When the inflaton field starts from a large field value and
then rolls down towards the minimum value of potential function, the
field value is about to vanish at this point. This is known as
chaotic inflation in which inflaton field is greater than $M_{Pl}$
and ends when inflaton field is nearly close to $M_{Pl}$. Models
which correspond to chaotic inflation are known as large field
models. To investigate such type of inflation, we have taken EoS
with a deviation parameter which describes quasi-de Sitter expansion
and leads to an elegant exit from inflation to deceleration phase.
We have furnished some basic features of inflation and formulated
Hubble flow functions as well as slow-roll parameters in fluid
cosmology and scalar field for a power-law model of $f(R)$ gravity.

We have analyzed inflation by taking different values of $\nu$ with
$\mu=1$ and $\mu\neq1$. We have calculated slow-roll parameters,
spectral index and tensor-scalar ratio for all these values. The
results can be summarized as follows.
\begin{itemize}
\item For $\nu=\frac{1}{6}$, we have constructed the graphical
analysis for $m=0.3,~0.5,~0.8$ and $n=0.9,~0.8,~0.75$. The
tensor-scalar ratio $r$ shows consistency with Planck observations
for $n=0.75$ with $m\geq-0.69 ,~m\geq-12.31$ and $m>0$ ($m\neq1,~2$)
whereas for $-0.7\leq{m}\leq-12.3$, there is no graphical
interpretation. For $n=1.98$, we have obtained consistent results
for $0.727\leq{m}\leq1.335$. The compatible number of e-folds are
$\mathcal{N}=54,~\mathcal{N}=41,~\mathcal{N}=34,~\mathcal{N}=29$ for
$n=0.7,~1.1,~1.5,~1.98$, respectively. We have found consistent
results as e-folds as well as anisotropic parameter are getting
smaller when $n$ is getting larger.
\item In case of $\nu=\frac{1}{3}$, the range of $n$
is $0.4\leq{n}\leq1.4,~0.4\leq{n}\leq1.3,~0.4\leq{n}\leq1.3$ for the
above values of anisotropic parameter. For $n=0.4,~0.8,~1.1$ and
$1.4$, the e-folds are found to be
$\mathcal{N}=148,~\mathcal{N}=74,~\mathcal{N}=53$ and
$\mathcal{N}=29$, respectively. The tensor-scalar ratio is
compatible with recent observational data for $n=0.4$ with all
values of $m$ instead of $-1\leq{m}\leq-5,~(m\neq0,~1,~2)$. When
$n=1.6$, the results are found to be consistent for $0<m\leq0.2$.
\item When $\nu=\frac{2}{3}$, the range of $n$
becomes $1.1\leq{n}\leq1.83,~1.1\leq{n}\leq1.92,~1.1\leq{n}\leq1.98$
for the same anisotropy values. The e-folds gives
$\mathcal{N}=78,~49,~34,~30$ for $n=1.1,~1.5,~1.83$ and $1.98$,
respectively. The tensor-scalar ratio turns out to be compatible
with Planck constraint for $n=1.1$ and $n=1.98$ with
$-0.42\leq{m}\leq29.4$ and $0.727\leq{m}\leq1.335$, respectively.
\end{itemize}

In fluid cosmology, we have calculated Hubble flow functions
$(\epsilon_1,~\epsilon_2)$ which recover expressions of the spectral
index and tensor-scalar ratio for the scalar field. We have also
evaluated the value of $f(R)$ which corresponds to Starobinsky
inflationary model for $n,~f_0,~m=1$. For $\mu\neq1$, we have taken
$\mu=2$ and constructed observational parameters for $\nu$ and
expressions found in fluid cosmology. We again recover these
observational parameters as well as Hubble flow functions. We have
investigated inflation with scalar field and developed expressions
of kinetic and potential functions. In this case, the tensor-scalar
ratio is incompatible to Planck constraints whereas $\mathcal{N}=59$
for all values of $n$ and $\nu$.

Myrzakulov et al. (2015) analyzed the dynamics of inflation via
scalar field as well as fluid cosmology through isotropic
homogeneous universe model in $f(R)$ gravity. To explore the
existence of inflationary epoch with smooth ending, they considered
quasi-de Sitter expansion with the EoS parameter evolving e-folds.
In the scalar field representation, they obtained quadratic form of
potential function compatible with massive scalar field. They also
claimed that for $\mu=2$, the amount of e-folds as well as
tensor-scalar ratio is found to be consistent with recent Planck's
constraints whereas this ratio appears to be larger for $\mu=1$. In
the presence of inhomogeneous fluids, they determined explicit
solutions which preserve the same behavior as scalar field to
produce inflation. In this paper, we have found consistent range of
e-folds as well as tensor-scalar ratio relative to different ranges
of anisotropic parameter and for all considered values of $\nu$
whereas in case of $\mu=2$, the tensor-scalar ratio exceeds from
Planck's suggested limit. We have also found expressions of kinetic
and potential energies for scalar field with
$\nu=\frac{1}{6},~\frac{1}{3}$ and $\frac{2}{3}$ which yield linear,
quadratic and quartic potential models, respectively. In case of
inhomogeneous fluid, the density dependent EoS parameter identifies
the same behavior of observational parameters as in the presence of
scalar field. It is worth mentioning here that all our results are
consistent with isotropic and homogeneous universe for $n,~m,~f_0=1$
(Myrzakulov et al. 2015).
\newpage
\vspace{0.25cm}

{\bf Acknowledgments}

\vspace{0.25cm}

This work has been supported by the \emph{Pakistan Academy of
Sciences Project}.

\end{document}